\documentclass[aps,prb,twocolumn,showpacs,superscriptaddress,preprintnumbers,amsmath,amssymb]{revtex4-1}

\usepackage{graphicx}
\usepackage{epsfig}
\usepackage{dcolumn}
\usepackage{bm}
\usepackage{longtable}
\usepackage{color}

\begin{document}


\title{Anomalous hysteresis as an evidence for a magnetic field-induced chiral superconducting state in LiFeAs}



\author{G.\ Li}
\affiliation{National High Magnetic Field Laboratory, Florida
State University, Tallahassee-FL 32310, USA}
\author{R.\ R.\ Urbano}
\affiliation{National High Magnetic Field Laboratory, Florida
State University, Tallahassee-FL 32310, USA}
\affiliation{Instituto de F\'{i}sica ``Gleb Wataghin", UNICAMP, 13083-859 Campinas, Brazil}
\author{P.\ Goswami}
\affiliation{National High Magnetic Field Laboratory, Florida
State University, Tallahassee-FL 32310, USA}
\author{C.\ Tarantini}
\affiliation{National High Magnetic Field Laboratory, Florida
State University, Tallahassee-FL 32310, USA}
\author{B.\ Lv}
\affiliation{Texas Center for Superconductivity, University of Houston, Houston, Texas 77204-5002, USA}
\author{P.\ Kuhns}
\affiliation{National High Magnetic Field Laboratory, Florida
State University, Tallahassee-FL 32310, USA}
\author{A.\ P.\ Reyes}
\affiliation{National High Magnetic Field Laboratory, Florida
State University, Tallahassee-FL 32310, USA}
\author{C.\ W.\ Chu}
\affiliation{Texas Center for Superconductivity, University of Houston, Houston, Texas 77204-5002, USA}
\author{L.\ Balicas} \email{balicas@magnet.fsu.edu}
\affiliation{National High Magnetic Field Laboratory, Florida
State University, Tallahassee-FL 32310, USA}


\date{\today}

\begin{abstract}
Magnetometry measurements in high quality LiFeAs single-crystals reveal a change in the sign of the magnetic hysteresis in the vicinity of the upper critical field $H_{c2}$, from a clear diamagnetic response dominated by the pinning of vortices, to a considerably smaller net hysteretic response of opposite sign, which \emph{disappears} at $H_{c2}$. If the diamagnetic response at high fields results from pinned vortices and associated screening super-currents, this sign change must result from currents circulating in the opposite sense, which give rise to a small field-dependent magnetic moment \emph{below} $H_{c2}$. This behavior seems to be extremely sensitive to the sample quality or stoichiometry, as we have observed it only in a few fresh crystals, which also display the de Haas van Alphen-effect. We provide arguments against the surface superconductivity, the flux compression, and the random $\pi$ junction scenarios, which have been previously put forward to explain a paramagnetic Meissner effect, below the lower critical field $H_{c1}$. The observed anomalous hysteresis at high fields will be compatible with the existence of chiral gap wave-functions, which possess a field dependent magnetic moment. Within a Landau-Ginzburg framework, we demonstrate how a $(d_{x^2 - y^2} + id_{xy})$ or a $(p_x+ip_y)$ chiral superconducting component can be stabilized in the mixed state of $s_{\pm}$ superconductor, due to the combined effects of the magnetic field and the presence of competing pairing channels. The realization of a particular chiral pairing depends on the microscopic details of the strengths of the competing pairing channels.
\end{abstract}

\pacs{74.70.Xa, 74.25.Dw, 74.62.Dh, 74.25.fc}

\maketitle

\section{Introduction}

LiFeAs \cite{tapp} is a stoichiometric compound belonging to the new family of layered iron pnictides, which displays a superconducting transition at a critical temperature $T_c \simeq 18$ K.
In contrast to most Fe based superconductors, LiFeAs becomes superconducting without doping a parent metallic antiferromagnet \cite{liu}.  Local Density Approximation (LDA) calculations indicate that the Fermi surfaces (FS) of virtually all non-magnetic Fe pnictide compounds consist of two cylindrical sheets of electron character at the $M$ point, and depending on the doping level, of at least two more cylinders of hole-character at the $\Gamma$ point of the first Brillouin zone (FBZ) \cite{singh}. In the simplest scenarios \cite{singh}, the antiferromagnetism is believed to emerge from a Peierls-like instability associated with nearly nested electron- and hole-like cylindrical Fermi-surfaces (FS). Therefore, from the perspective of band structure calculations, LiFeAs should display itinerant antiferromagnetism instead of a superconducting ground state \cite{singh}. This tendency towards antiferromagnetism produces antiferromagnetic fluctuations, which have been claimed to be responsible for the superconducting pairing \cite{mazin}. Although, for these multi-orbital systems characterized by a strong Hund's coupling, alternative pairing scenarios including triplet pairing, have been proposed \cite{Dai,lee,Raghu,Goswami1,Goswami2}.

Initial angle resolved photoemission spectroscopy (ARPES) measurements on LiFeAs have indicated very different relative sizes for the electron and hole-like FSs, implying the absence of FS nesting and a concomitant itinerant magnetism \cite{borisenko}. This same study in contrast to Ref. \onlinecite{inosov}, but in agreement with more recent measurements \cite{umezawa}, finds that the superconductivity in LiFeAs is multi-band in nature with gaps in both sets of FSs. This conclusion is supported by penetration depth measurements \cite{song, kim, hashimoto}, which find at least two $s$-wave gaps, whose relative amplitude is within a factor of two. The additional support to the results of Ref. \onlinecite{inosov} is provided by the heat-capacity \cite{wei} and the nuclear magnetic resonance studies, which find the relative amplitude of two gaps to be within a factor of three \cite{li,jrglic}. The $s \pm$ state \cite{mazin} proposed for 122 iron pnictides, whose corresponding gap wave-function has nodes located in between the cylindrical FSs is in agreement with a multi-band superconducting state with gaps on both sets of FSs.

The above results are in sharp contrast with a recent point contact spectroscopy measurement, which finds no support for elementary singlet-pairing symmetries. Instead, a chiral $(p_x+ip_y)$-wave symmetry pairing has been claimed to provide the best fit to their quasi-particle interference patterns \cite{hanke}. Moreover, a nuclear magnetic resonance (NMR) and a nuclear quadrupole resonance (NQR) studies in \emph{some crystals}, have found a constant Knight shift and an upturn in the NQR relaxation rate across $T_c$ for fields along the $ab$-plane, which is unexpected for conventional superconductivity \cite{baek}.

On the other hand, the application of high magnetic fields has lead to distinct phase diagrams \cite{zhang, khim, cho, kurita}, with the values of the superconducting upper-critical fields extrapolated to zero temperature $(H_{c2}(T \rightarrow 0 K))$ varying by more than 30 \% \cite{zhang, khim, cho,kurita}. This suggests that the superconductivity in LiFeAs is particularly sensitive to impurities or variations in stoichiometry. Remarkably, the superconducting phase-boundary at low $T$s for in-plane fields has been claimed \cite{cho} to show evidence for the Fulde-Ferrel-Larkin-Ovchinnikov (FFLO) state \cite{burkhardt}. However, for fields applied along the c-axis, the thermal conductivity measurements have not found any evidence for a field-induced phase-transition \cite{tanatar}. Therefore, we do not have a consensus regarding the nature of the pairing and the phase diagram in the entire $H-T$ plane.

In order to clarify i) the potential existence of additional phases within the superconducting phase diagram of LiFeAs and ii) the geometry of the FS through the de  Haas van Alphen-effect, we have performed magnetic torque and  magnetization measurements at high fields in high quality single crystals of LiFeAs, with a middle point superconducting transition temperature $T_c \simeq 17.1$ K. The quality of our samples is indicated by the observation of the de Haas-van-Alphen-effect, and also by the $^{75}$As NMR spectrum, indicating the absence of magnetism or inhomogeneities. The de Haas van Alphen results indicate that the cross sectional areas of the hole-like FSs may be considerably smaller than the values predicted by the band structure calculations, explaining the absence of antiferromagnetism. In particular, we observe a reversal in the sign of the magnetic hysteresis, from a diamagnetic to a much smaller but ``paramagnetic"-like response, \emph{within the superconducting state} of LiFeAs at high fields, which disappears at $H_{c2}$. We clarify however, that the term ``paramagnetic" is never applied to the net hysteresis but to the conventional magnetic response of a given system. However, and in absence of a proper term, we will use it here to contrast this anomalous hysteresis with the diamagnetic-like hysteretic response seen in a type-II superconducting state. Such an effect has never been reported for any physical system. The concomitant disappearance of the anomalous hysteresis with the destruction of the superconductivity at $H_{c2}$, demonstrates that the observed phenomenon is strictly associated with the underlying paired state.  In addition, the paramagnetic response is observed only above the lower critical field $H_{c1}$, as opposed to the ``paramagnetic Meissner effect" (also known as the Wohlleben-effect) observed below $H_{c1}$. Therefore the mechanisms such as flux compression \cite{geim,moshchalkov,koshelev}, random $\pi$ junctions \cite{sigrist} etc.,
which have been previously proposed to explain the Wohlleben-effect, can not be responsible for the observed anomalous hysteresis at high magnetic fields.

Our observation will rather be consistent with a superconducting state, which possesses a field dependent orbital magnetic moment in the mixed state. Based on a Landau-Ginzburg analysis, we argue that a chiral superconducting component of $(d_{x^2 - y^2} + id_{xy})$ or $(p_x+ip_y)$ symmetry can be stabilized inside the mixed state of a $s_{\pm}$ superconductor, due to the combined effects of magnetic field and the presence of competing pairing channels. The stabilized chiral components are intimately tied to the presence of vortex solutions of the $s_{\pm}$ state, and consequently the orbital magnetic moment arising from the chiral components can lead to the anomalous hysteretic response.

\section{Experimental Results}
LiFeAs single-crystals were synthesized by using LiAs flux method. Bulk superconductivity was observed in crystals through heat capacity measurements with an onset at $\sim 16$ K. Torque measurements were performed by using a capacitive cantilever beam configuration. Resistance measurements were performed by using a four terminal configuration. All samples were cleaved to expose fresh shiny surfaces and all crystals were cut with a razor blade to have a typical dimension of ~ 1.2 x 1 x 0.75 mm$^3$. To make contacts samples were kept under Ar atmosphere and on a hot plate at $120^{\circ}$ C, where Au wires were attached with silver epoxy.  Torquemeters were placed on a single-axis rotator inserted into either a $^3$He cryostat coupled to a superconducting magnet or a $^3$He cryostat coupled to the 45 T hybrid-magnet. Good thermalization with the temperature of the liquid helium bath was achieved by using large amounts of $^3$He as exchange gas or $^3$He in its liquid state. The angle was measured with Hall probes. Magnetization was measured in a vibrating sample magnetometer, coupled to a superconducting magnet. The same crystals used for the magnetometry measurements were mounted on a NMR probe equipped with a single axis goniometer, which allowed a fine alignment of the crystallographic axis with the external field.

\begin{figure}[htbp]
\begin{center}
\epsfig{file=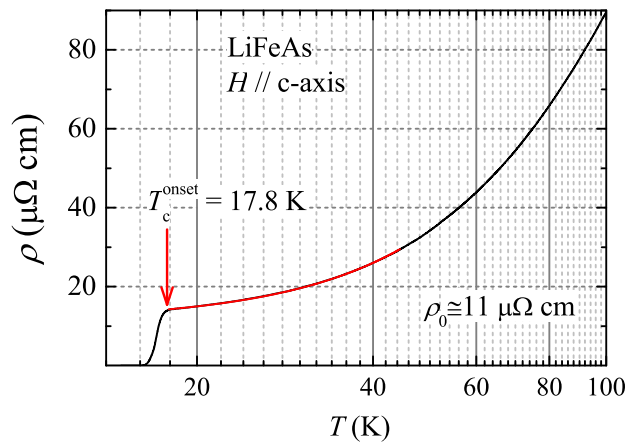, width = 8.2 cm}
\caption{(color online)  Resistivity ($\rho$) as a function of the temperature for one of our LiFeAs single crystals at zero field.
$T_c^{\text{onset}} =17.8$ K indicates the onset of the superconducting transition. Red line is a fit to $\rho = \rho_0 + AT^2$ from which one extracts the value $\rho_0 \simeq 11$ $\mu\Omega$ cm.}
\end{center}
\end{figure}
Figure 1 shows the typical resistivity of one of our single crystals as a function temperature ($T$) under zero field ($H$). The onset of the superconducting transition is observed at $T_c^{\text{onset}} = 17.8$ K with a transition width $\Delta T_c \simeq 0.8$ K. $\Delta T_c$ is defined as $T(\rho (90 \%)) - T(\rho (10 \%))$, where $T(\rho (90 \%))$ corresponds to the value in temperature where the resistivity reaches 90 \% of $\rho_n$ or its value in the metallic state just above the resistive transition, while $T(\rho (10 \%))$ corresponds to the temperature where the resistivity reaches 10 \% of $\rho_n$. Red line is is a fit to $\rho = \rho_0 + AT^2$ from which one extracts a residual resistivity of $\rho_0 \simeq 11$ ($\mu\Omega$ cm). Notice that given the size of the contacts relative to the sample size the error bars in the determination of  $\rho_0$ could be as large as 50 \%. Therefore LiFeAs displays the smallest residual resistivity by at least one order of magnitude, among the pure As-based Fe-pnictide superconductors.

\begin{figure}[htbp]
\begin{center}
\epsfig{file=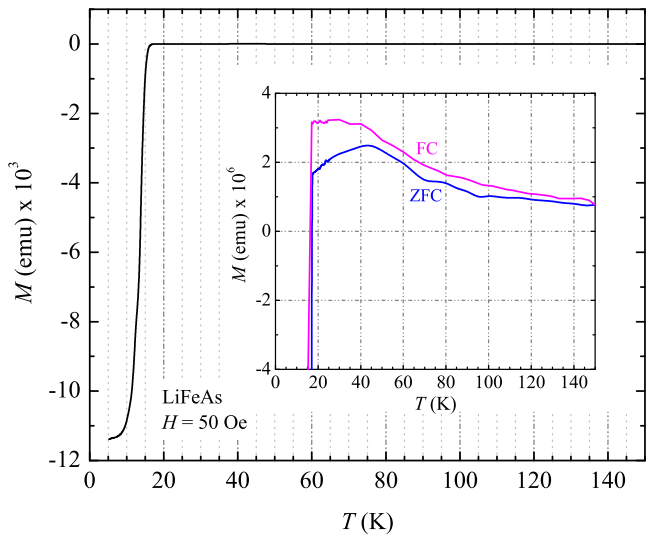, width = 8.5 cm}
\caption{(color online) Main panel: Magnetization ($M$) as a function of the temperature for one of our LiFeAs single crystals acquired under a field of 50 Oe and after cooling the sample at zero field. Inset: $M$ as a function of temperature but in an amplified scale showing the magnetic response of the metallic state under zero-field- (blue line) and under field-cooled (magenta line) conditions.}
\end{center}
\end{figure}
Fig. 2 shows the magnetization $M$ as a function of the temperature for one of our LiFeAs single crystals. The inset displays $M$ as a function of temperature but in an amplified scale showing the magnetic response of the metallic state under zero-field- (blue line) and under field-cooled (magenta line) conditions. The apparent hysteresis is an experimental artifact, since the magnetic response of the metallic state is comparable in magnitude to the sensitivity of the instrument. As seen, the magnetic response of the metallic state is several orders of magnitude smaller than the characteristic diamagnetic one from the superconducting state. This by itself, points towards the absence of localized Fe moments which would provide a sizeable contribution, i.e. comparable in magnitude to the diamagnetic one. This is further confirmed by the temperature dependence of the metallic state magnetization, which does not display the characteristic Curie-Weiss susceptibility expectable for localized magnetic moments. Therefore,  $M$ confirms the absence of both localized magnetic impurities and of long-range magnetic-order in our LiFeAs single-crystals.

\begin{figure}[htbp]
\begin{center}
\epsfig{file=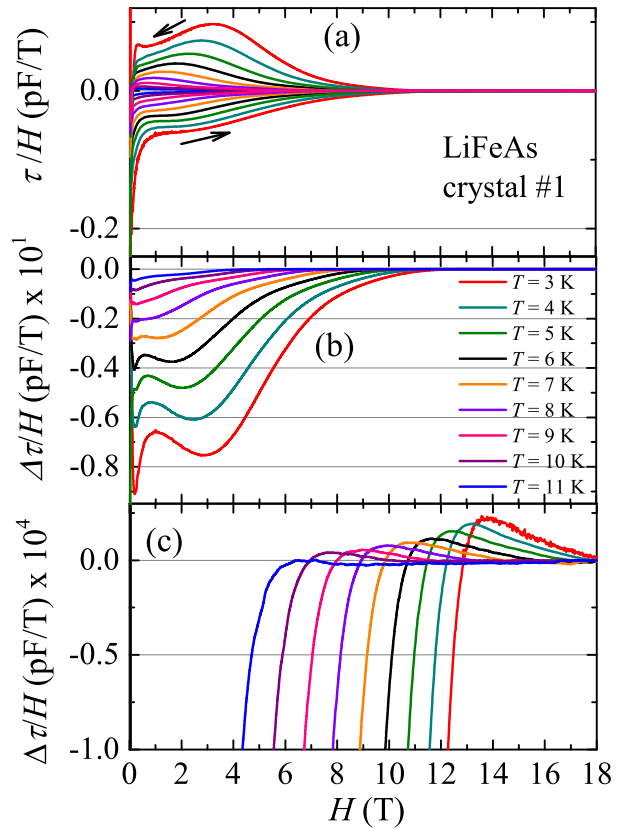, width = 8 cm}
\caption{(color online) (a) Magnetic torque $\tau$ as a function of the magnetic field $H$ applied along a direction nearly perpendicular to the inter-layer c-axis, and normalized by $H$ for a LiFeAs single crystal (crystal $\sharp 1$) at several temperatures $T$. (b) $\Delta \tau /H = (\tau_{H_\text{inc}}/H_\text{inc} - \tau_{H_\text{dec}}/H_\text{dec})/2$ or the pure hysteretic and diamagnetic response in the magnetic torque. (c) $\Delta \tau /H$ in an amplified scale.  Notice how the diamagnetic signal is followed by an anomalous positive (as if paramagnetic-like) hysteresis at higher fields which grows as $T$ is lowered but is suppressed as $H \rightarrow H_{c2}$.}
\end{center}
\end{figure}

\begin{figure}[htp]
\begin{center}
\epsfig{file=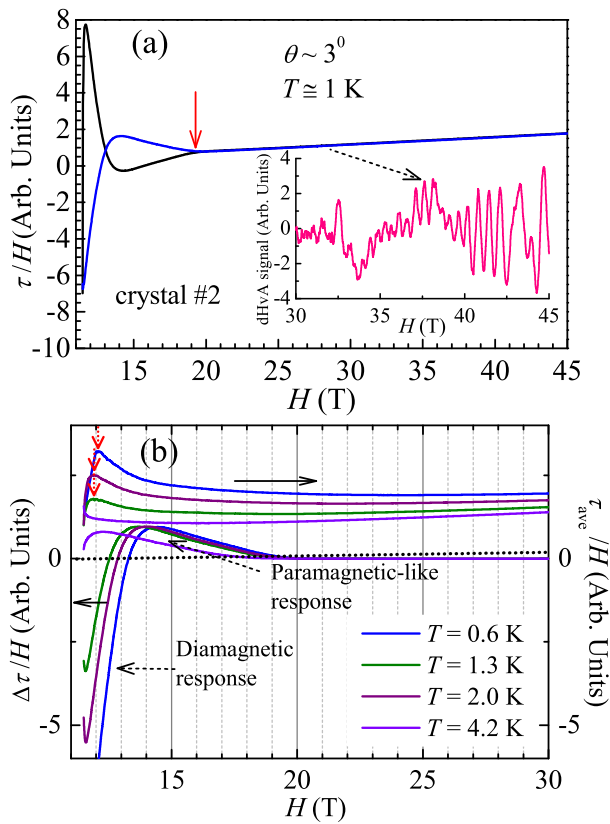, width = 8 cm}
\caption{(color online)  (a) $\tau/H$ as a function of $H$, for fields between 11.5 to 45 T, at $T = 1$ K and for an angle $\theta = 3^{\circ}$ between $H$ and inter-planar c-axis, for a \emph{second} LiFeAs single-crystal ($\sharp 2$). Notice the presence of a second rather symmetric hysteresis loop whose sign is opposite to the diamagnetic one observed at lower fields. Once this second loop closes one observes the expected linear in $H$ dependence for the paramagnetic metallic state.  Once this term is subtracted by fitting it to a straight line, an oscillatory component is observed, i.e. the de Haas van Alphen-effect (shown in the inset). (b) Respectively,  $\Delta \tau /H $  (left vertical axis) and the average $ \tau_{\text{ave}} /H$ between both field-increasing and decreasing torque traces (right vertical axis), for a few temperatures. Curves are vertically displaced for clarity. This average is proportional to the reversible linear component in $\tau/H$ except at lower fields where one observes an anomaly (red vertical arrows) which becomes more pronounced as $T$ is lowered. This anomaly is observed well inside the diamagnetic hysteretic region suggesting a possible phase transition.}
\end{center}
\end{figure}

The upper panel of Fig. 3 (a) shows the magnetic torque  $\overrightarrow{\tau} = \mu_0 \overrightarrow{M} \times \overrightarrow{H}$, where $\overrightarrow{M}$ is the bulk magnetization, for a LiFeAs single crystal as a function of the field $H$ applied nearly parallel to the inter-layer direction and for several temperatures $T$.  For a layered metallic system as LiFeAs, one can readily demonstrate that $\tau = \mu_0/2 (\chi_{aa} - \chi_{zz})H^2 \sin 2 \theta$  where  $\chi_{aa}$ and $\chi_{zz}$ are the in-plane and the out of plane components of the susceptibility tensor, respectively. In Fig. 3 (a) for each temperature both the increasing ($H_\text{inc}$) and decreasing ($H_\text{dec}$) field sweeps are included. One observes a pronounced hysteresis loop $\Delta \tau = (\tau_{H_\text{inc}} - \tau_{H_\text{dec}})/2$ between increasing and decreasing field sweeps (indicated by the arrows), which according to the standard Bean model \cite{bean} is proportional to the superconducting critical current $J_c$, or $\Delta \tau \propto \Delta M \propto J_c = F_p/ \mu_0 H$ where $F_p$ is the vortex pinning force density. The vortex pinning mechanisms in LiFeAs have already been studied by other groups \cite{pramanik} and it is not the focus of this manuscript. Fig. 3 (b) shows  $\Delta M \propto \Delta \tau /H = (\tau_{H_\text{inc}}/H_\text{inc} - \tau_{H_\text{dec}}/H_\text{dec})/2$ or the pure hysteretic and diamagnetic response in the magnetic torque. Fig. 3 (c) shows $\Delta \tau /H $ in an amplified scale. As seen $\Delta \tau /H $ displays a surprising change in the sign of the magnetic hysteresis/irreversibility, from diamagnetic to an anomalous hysteretic response having the opposite sign.  This rather small ``paramagnetic" hysteresis grows as the temperature is lowered and progressively disappears as the field increases in contrast to what is expected for a field-induced magnetic state, indicating that it is intrinsically associated to the superconducting order parameter.  It is easy to demonstrate that the sign of the pure hysteretic response $\Delta M$ between field-up and -down sweeps in ferromagnets and in superconducting materials has the exact same sign. Therefore, this anomaly cannot be attributed to a conventional re-arrangement among magnetic domain walls. Furthermore, We changed the magnet sweep rate from 0.5 to 0.05 T/min, obtaining exactly the same curve(s).

This remarkable hysteresis is more evidently exposed in Fig. 4 (a) which shows $\tau/H$ for a \emph{second} single-crystal (crystal $ \sharp 2$) as a function $H$ and for fields between 11.5 and 45 T at $T = 1$ K and an angle $\theta = 3^{\circ}$ between $H$ and the c-axis of the crystal. At $H \sim 13$ T the diamagnetic hysteresis loop closes and a second loop of opposite sign emerges. When this second loop closes one observes a linear in $H$ dependence as expectable for a metallic paramagnetic state, which can be subtracted revealing an oscillatory component, i.e. the de Haas van Alphen effect (dHvA).  A detailed dHvA study in LiFeAs is not among the main objectives of this communication, nevertheless we advance below a few implications of our observations. Fig. 4 (b) shows both the average $\tau_{\text{ave}}/H$ between field-increasing and -decreasing $\tau/H$ branches as well as $\Delta \tau /H$ for several temperatures. $\Delta \tau H$ behaves similarly as seen in Fig. 3, while $\tau_{\text{ave}}$ which at larger fields (where the hysteresis becomes smaller) is proportional to the reversible or non-hysteretic magnetic response displays at lower fields an anomaly (indicated by red arrows) that sharpens as the temperature is lowered. This anomaly indicates a possible phase-transition within the diamagnetic hysteretic region just before the emergence of the anomalous ``paramagnetic" hysteresis.
\begin{figure}[htp]
\begin{center}
\epsfig{file=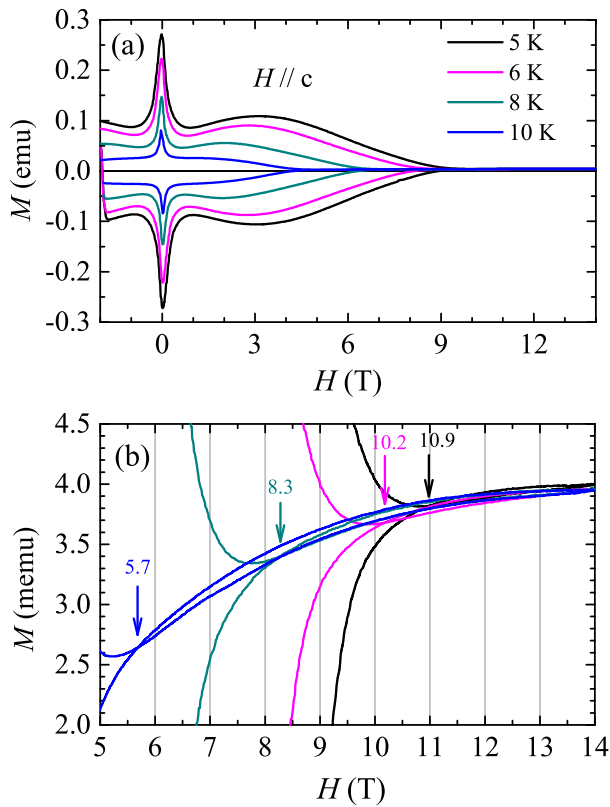, width = 8.0 cm}
\caption{(color online) (a) Magnetization $M$ as a function of $H$ applied along the inter-planar direction for a few temperatures, and respectively for increasing and decreasing field sweeps. (b) Same as in (a) but in a limited field and magnetization range. This data was acquired by using a vibrating sample magnetometer. Notice that the magnetization $M =  \chi_{zz} H$ branches, similarly to the torque data, also cross at higher fields. }
\end{center}
\end{figure}

The above observations contrast markedly with torque results show in Ref. \onlinecite{kurita}, which reveal far more asymmetric hysteresis loops with apparently, no change in the sign as seen here.
The hysteresis loops in Ref. \onlinecite{kurita} close at considerably lower fields than the ones reported here, thus indicating that our samples display higher upper critical fields.
Therefore, we conclude that our samples should be of higher quality at the moment of performing the torque measurements, which is likely to explain this difference in behavior.

\begin{figure}[htp]
\begin{center}
\epsfig{file=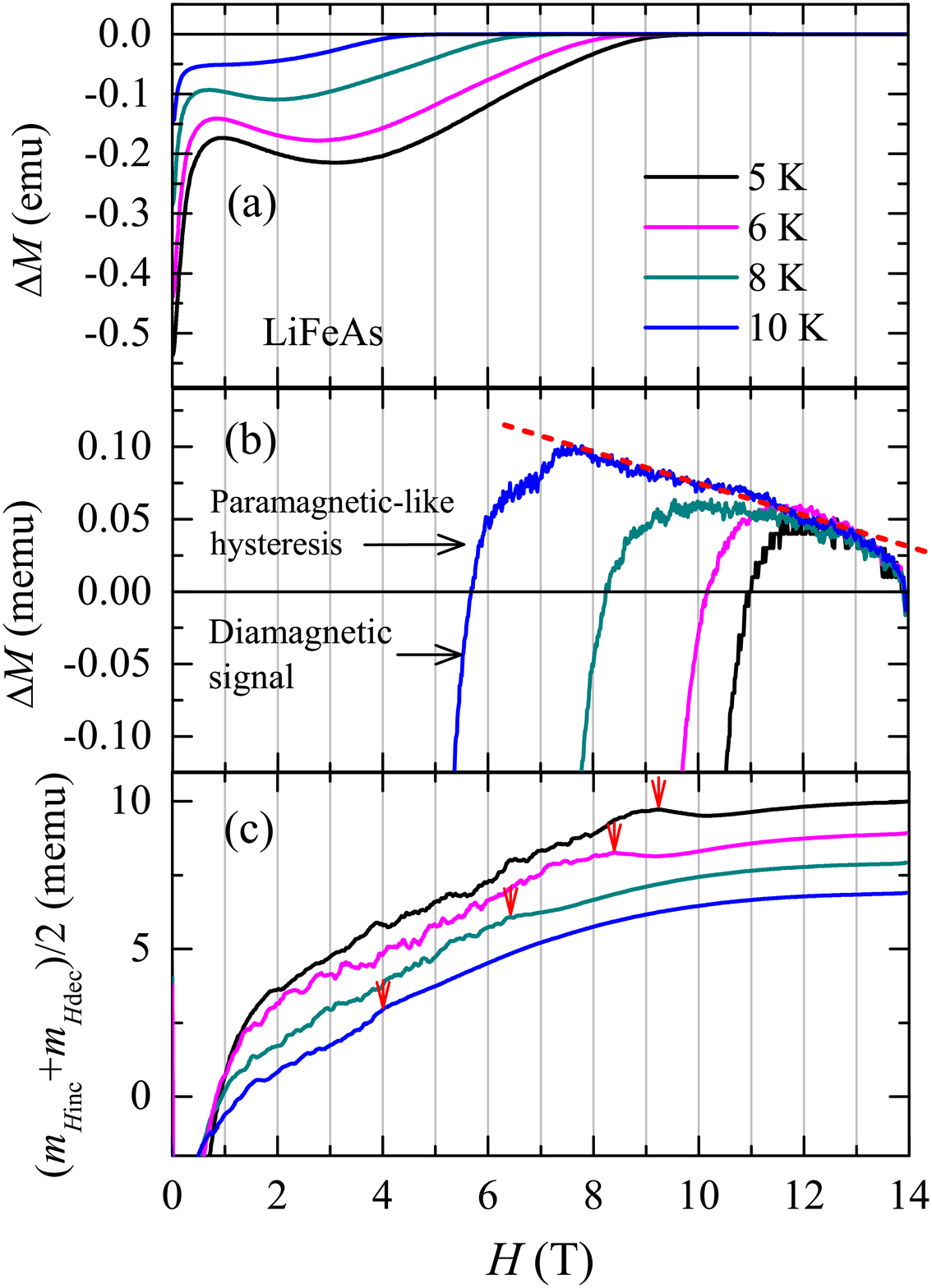, width = 8.0 cm}
\caption{(color online) (a) Pure irreversible component in the magnetization as the difference $\Delta M$ between the field increasing and decreasing traces. (b) Same as in (a) but in an amplified scale, showing the crossover from the diamagnetic to a paramagnetic-like irreversibility. Notice that this paramagnetic response is more than 3 orders of magnitude smaller than the diamagnetic one and becomes progressively smaller as the field is increased. (c) Average between both magnetization traces which is proportional to the reversible component in the magnetization. This so-defined reversible component increases with field and nearly saturates at the highest fields, showing a mild anomaly (indicated by red arrows) before saturating.}
\end{center}
\end{figure}
Since the magnetic torque in layered materials is proportional to the anisotropy of the magnetic susceptibility, the behavior shown above could be attributed to some dramatic reconfiguration of the vortex matter, which might lead to a reversal in the relative size of the terms in the susceptibility tensor thus changing the sign of the torque ($\tau \propto (\chi_{aa} - \chi_{zz})$). To study this possibility we performed magnetization $M =  \chi_{zz} H$  measurements in the same single crystals. Fig. 5 (a)  displays $M$ as a function of $H$ for a few
temperatures with Fig. 5 (b) showing $M$ as a function of $H$ in an amplified vertical scale. Notice, i) the very strong similarities between $M$ and $\tau/H$ and ii) how the hysteresis branches again cross at higher fields.

Figure 6 (a) shows the net hysteretic response $\Delta M = (M(H_{\text{inc}})-M(H_{\text{dec}}))$ and Fig. 6 (b) displays $\Delta M$ in an amplified vertical scale. The important observation is that $M$, \emph{measured through a quite distinct experimental technique} or vibrating sample magnetometry, also shows the anomalous paramagnetic hysteresis which is at least 1000 times smaller than the corresponding diamagnetic one, in agreement with the magnetic torque previously shown.  Notice also that this signal disappears as the field increases on approaching $H_{c2}$ (as indicated by the red dashed line), implying again that it \emph{cannot} be associated to localized moments or magnetic domains (a net magnetic signal is expected to \emph{increase} with field). $M$ clearly indicates that this anomalous hysteretic response cannot be attributed to any relative change in the components of the susceptibility tensor. Finally, Fig. 6 (b) shows the average magnetization or $M_{\text{av}} = (M(H_{\text{inc}})+M(H_{\text{dec}}))$ as a function of field. Curiously, $M_{\text{av}}$ which corresponds to a very small background magnetization increases with field saturating at the largest values, as expectable for a net magnetic signal (although within the superconducting state). Similarly to the magnetic torque, very mild anomalies (indicated by the red arrows) are also seen in $M_{\text{av}}$ at fields where $\Delta M$ is well within the diamagnetic regime. At this point, we can conclude that there is a remarkable agreement between two very different experimental techniques probing the magnetic response of our LiFeAs single crystals.

\begin{figure}[htp]
\begin{center}
\epsfig{file=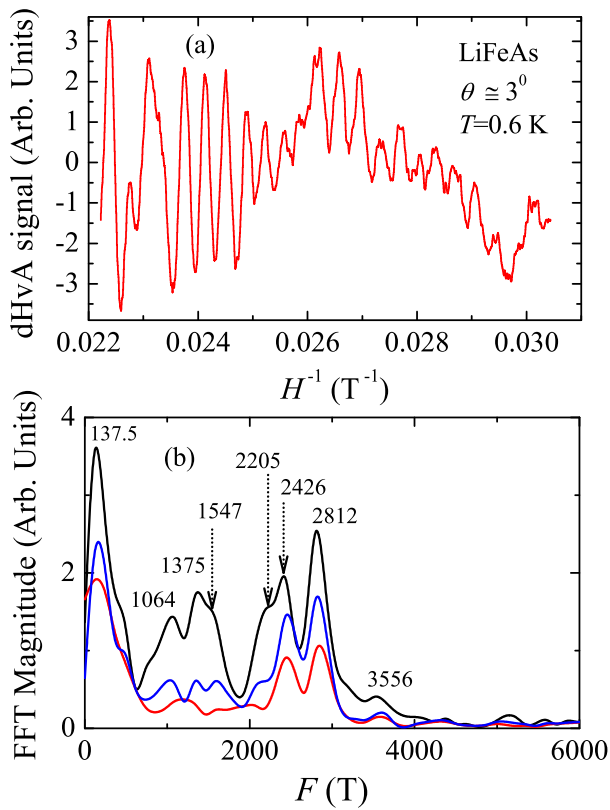, width = 8 cm}
\caption{(color online) (a) Oscillatory component, or the de Haas van Alphen effect superimposed onto the torque signal of a LiFeAs single crystal within its metallic phase as a function of the inverse field $H^{-1}$, at a temperature $T = 0.6$ K and an angle $ \theta = 3^{\circ}$ between the magnetic field and the inter-planar c-axis. (b) Magnitude of the fast Fourier transform of the oscillatory signal shown in (a) as a function of the cyclotronic frequency, showing at least 4 clear peaks and indications for several additional frequencies. The different traces correspond to distinct window filters.}
\end{center}
\end{figure}
\begin{figure}[htp]
\begin{center}
\epsfig{file=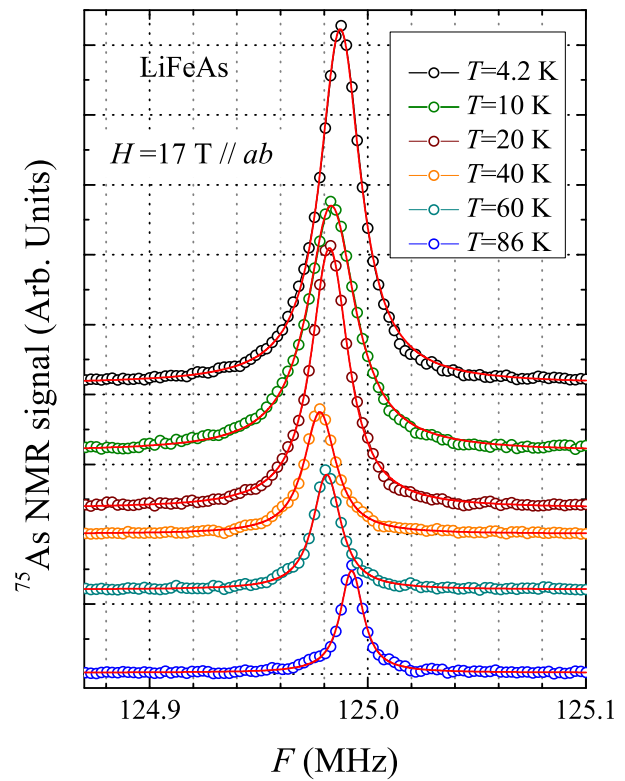, width = 8 cm}
\caption{(color online) Temperature dependence of the $^{75}$As NMR spectra for the same LiFeAs single-crystal with the magnetic field applied along the planar direction ($H \| ab$ = 17 T). The red solid lines are fits to a single Lorentzian line. The high quality of the fit and the quite narrow line-width (FWHM ~ 22.64 kHz at 4.2 K) observed for our LiFeAs single crystal discard the presence of either short or long range magnetic order. The of narrow Lorentzian NMR lines confirms the extreme high quality of our crystals and implies the absence of magnetic impurities or extrinsic phases.}
\end{center}
\end{figure}
To expose the quality of our crystals, we show in Fig. 7 (a)  the oscillatory component superimposed into our torque signal as a function of $H^{-1}$ (de Haas van Alphen signal) in the same single crystals used for torque and magnetization measurements. The Fast Fourier transform (FFT) of the oscillatory signal is shown in Fig. 7 (b). This figure demonstrates the quality of our crystals which as previously stated show no clear evidence for either impurities or magnetism. In effect, under a magnetic field, the crossing of the quantized electronic orbits (Landau levels) through the Fermi level produces oscillatory components in \emph{M}, or de the de Haas van Alphen (dHvA)-effect, which are periodic in $H^{-1}$ and whose fundamental frequencies $F$ are directly related to the extremal cross-sectional areas $A$ of the FS perpendicular to $H$ through the Onsager relation: $F = A (\hbar/2 \pi e)$. To observe dHvA oscillations the system must satisfy  $\omega_c \tau \gg 1$, where  $\omega_c$ is the cyclotron frequency and is the quasiparticle time of flight, i.e. the system must be clean. Band structure calculations predict at least 10 dHvA frequencies for LiFeAs \cite{putzke}. The fast Fourier transform (FFT) of the dHvA signal yields at least 4 frequencies, $F_1 = (150 \pm 15)$ T, $F_2 = (2450 \pm 25)$ T, $F_3 = (2840 \pm 30)$ T, $F_4 = (3550 \pm 50)$ T  with indications for a few more, in order words we detect a few more frequencies than those reported in Ref. \onlinecite{putzke} which were all attributed to cross sections from the electron-like FSs. In Fig. 3 (b) we show the FFT obtained by using three filters respectively, a rectangular window (black line), a Welch window (in blue) and a Blackman window (in red). The first two traces suggest three additional frequencies at $F_5 = (1060 \pm 40)$ T, $F_6 = (1375 \pm 25)$ T, and $F_7 = (1550 \pm 50)$ T, while the latter only reveals some spectral weight around 1200 T (the average frequency between $F_5$ and $F_6$).  These frequencies can only be attributed to hole-like FSs which would be at odds with the calculations shown in Ref. \onlinecite{putzke}, suggesting that the hole-like FSs would be smaller than the predicted ones, explaining the absence of nesting and consequently of itinerant magnetism. The observation of a large portion of the FS of LiFeAs, is a strong indication for the high quality of our single crystals.

Within a local (microscopic) point of view, one of the most efficient ways to investigate the origin of electronic and magnetic states and their homogeneity is through the analysis of the hyperfine interactions probed in a Nuclear Magnetic Resonance experiment.  In Fig. 8 we present the temperature dependence of the $^{75}$As NMR spectra with the magnetic field applied along the planar direction $(H \parallel ab = 17$ T).  The $^{75}$As NMR spectra taken at very high fields, which are similar to those shown in Ref. \onlinecite{li}, can be fitted to a single Lorentzian line shape, indicating that our sample is rather homogeneous and shows no traces of magnetic instabilities, crystallographic defects, distortions or strain, which would lead to additional broad lines. Therefore, all the evidence found by us, indicates homogeneous samples. Nevertheless, we found that this anomalous hysteresis disappears as a function of time, after the Apiezon grease covered single-crystals were briefly exposed to air, which is known to degrade their quality thus indicating that this behavior is quite sensitive to sample degradation. In fact, the suppression of this anomalous state might explain the differences among the superconducting diagrams in Refs. \onlinecite{zhang, khim, cho, kurita} since it would lead to lower upper critical fields.
\begin{figure}[htp]
\begin{center}
\epsfig{file=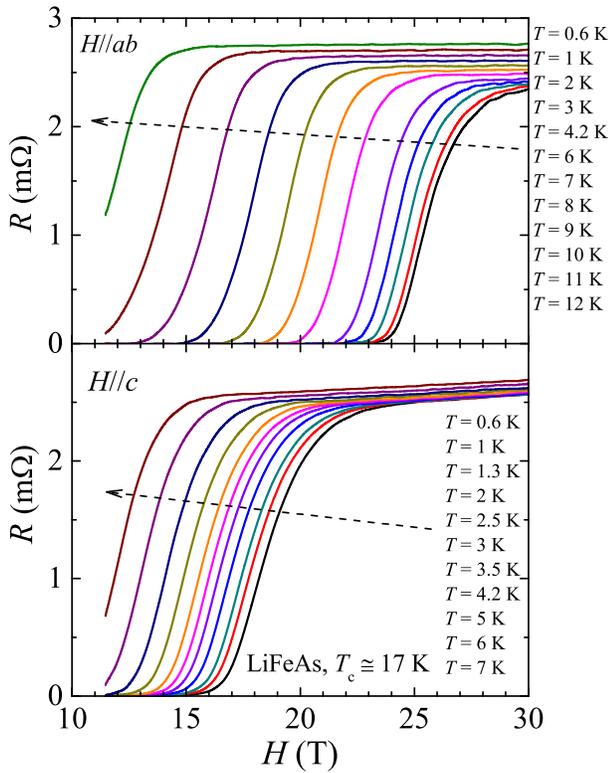, width = 8 cm}
\caption{(color online) (a) Resistance for a LiFeAs single crystal as a function of $H$ applied along a planar direction and for several temperatures. Here, the electrical current flows along a planar direction perpendicular to the external field. (b) Same as in (a) but for fields applied along the inter-planar direction. }
\end{center}
\end{figure}
\begin{figure}[htp]
\begin{center}
\epsfig{file=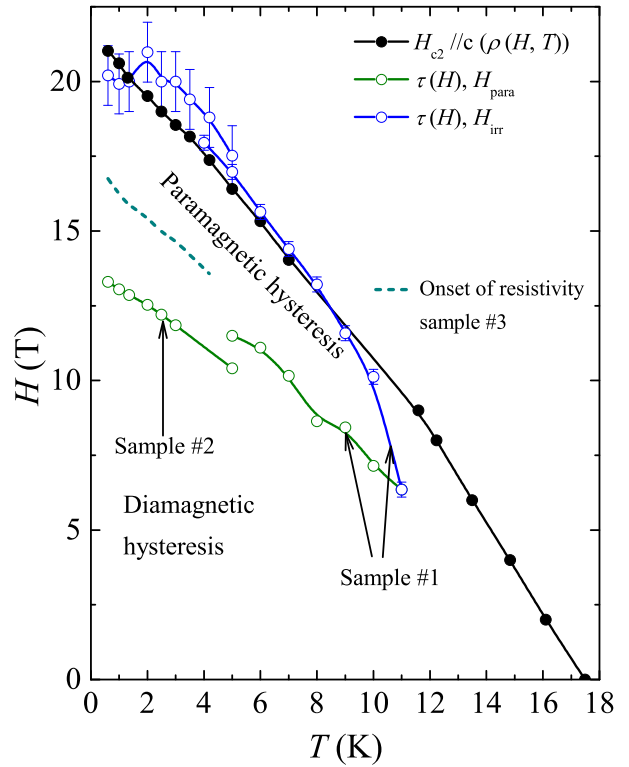, width = 8 cm}
\caption{(color online) From the data in Figs. 1, 4(a) and 4(b) superconducting phase diagram for LiFeAs for fields along the inter-planar direction as resulting from the present study: green markers connected by green lines depict the boundary between the diamagnetic and the paramagnetic irreversible behaviors, while the blue lines correspond to the irreversibility field $H_{\text{irr}}$ or the onset of the paramagnetic \emph{reversible} behavior. Black markers depict the onset of the superconducting transition for fields along the c-axis, i.e. the point in temperature where the resistance reaches 90 \% of its value in the normal state just prior to the transition (sample $\sharp$3). Darker dashed green line depicts the foot of the resistive transition, i.e. the point in temperature where the resistance emerges from its zero value (sample $\sharp$3 in (a)).}
\end{center}
\end{figure}

Figures 9 (a) and (b) show the resistance \emph{R} as function of \emph{H} applied along two orientations, i.e. a planar direction and the inter-planar axis, for a LiFeAs single crystal and for several temperatures, respectively. At first glance, the width of the transition would seem to be somewhat wide. However, we are still in the early stages of understanding the vortex pinning mechanisms in the Fe pnictides and most likely the width of the transition is intrinsic to the vortex physics in this material, instead of resulting from lower sample quality. For example, we found that F doping is a far more effective in pinning vortices than doping LaFeAsO\cite{li2} with Co which leads to a state akin to the vortex-liquid phase which is difficult to understand if one considers the fact that Co doping introduces site disorder into the FeAs layers. In the Fe$_{1+y}$Te$_{1-x}$Se$_x$ system on the other hand, one sees the superconducting transition to sharpen considerably as the temperature is lowered, or as the superconducting fluctuations are suppressed, although this system is quite disordered \cite{tesfay}. Finally, in the SmFeAsO$_{1-x}$F$_x$ system, which has a moderate superconducting anisotropy, one observes a marked change in pinning regime for fields applied along the \emph{ab}-plane and as the temperature is decreased below $T_c$: from a regime characterized by pinned Abrikosov vortices to a regime characterized by the intrinsic pinning of Josephson vortices as the coherence length shrinks below the inter-layer spacing upon cooling \cite{moll2}. These three examples illustrate the uniqueness of the vortex physics in the Fe pnictides, and therefore one should be cautious when analyzing the sample quality through the width of their superconducting transition under field.

The data in Fig. 9 is used to determine $H_{c2}(T)$ as shown in Fig. 10, which includes data only for fields oriented along the \emph{c}-axis, although we have observed the anomalous hysteretic response also for fields close to the \emph{ab}-plane. The onset of the resistive transition or the field for which the resistance reaches 90 \% of its value in the metallic state is the criteria used to extract $H_{c2}(T)$ (black markers).  In the same figure we have included the fields $H_{\text{para}}$ where the diamagnetic hysteresis is seen to cross towards a positive response (clear blue markers), and the irreversibility fields $H_{\text{irr}}$ for which the hysteresis completely disappears (blue markers). $H_{\text{irr}}$ coincides remarkably well with the onset of the superconducting transition as seen through the resistivity, although both measurements were performed in different crystals and the attachment of electrical contacts is expected to deteriorate the quality of the single-crystals (Li reacts with organic solvents). As seen in the figure, the superconducting region characterized by the paramagnetic irreversibility only emerges below 11 K, occupying a large portion of the phase diagram at lower temperatures and higher fields suggesting that it is not a property of the superconducting state at zero-field but likely a field-induced state.

Throughout this study we measured 7 crystals, 3 fresh crystals displayed the anomalous response reported here, and 4 other crystals from the same batch showing similar residual resistivities and the same middle point $T_c$ of $17.1$ K did not display neither the anomalous response nor the de Haas van Alphen effect. This points towards either very small variations in stoichiometry among crystals or/and to the critical role played by sample quality.

\section{Discussion}

There are various explanations for a paramagnetic Meissner response, which is seen below the lower critical field $H_{c1}$,  \cite{geim,moshchalkov,koshelev,sigrist}. These explanations either rely on the small system-size or the granularity of the sample. For a very small superconducting sample placed in the weak magnetic field, the magnetic length of the Cooper pair can be comparable to the size of the system. In this case, consideration of surface superconductivity and associated giant vortex state becomes important \cite{geim,moshchalkov}. The giant vortex state causes flux compression and leads to a paramagnetic Meissner effect in the low field limit. For our experimental set up, the magnetic length of the Cooper pair at large magnetic fields is much smaller than the system size. Therefore, the quantization of the orbital motion is relevant, which leads to the presence of multiple vortices, rather than a giant vortex state. The flux compression can also occur due to surface superconductivity, which is induced by inhomogeneous cooling \cite{koshelev}. However, the calculated magnetic moment due to the flux compression becomes paramagnetic only in the very low field limit. A diamagnetic moment is recovered in the high-field one, because this scenario is only applicable to the region of fields $H < H_{c1}$. In effect, for fields above this value the superconducting state is suppressed from the edges of the sample along a distance $x$ determined by the London penetration depth $\lambda_L$: the field $B(x)$ inside a superconductor is given by $B(x)= H_0 \exp(-x/\lambda_L)$ where $H_0$ is the external field. Therefore, strictly speaking, above $H_{c1}$ the field at the surface of the sample (or at $(x=0)$) is the external field $H_0$. Consequently, superconductivity should have been suppressed from the surface of the sample and this result is correct even in the presence of inhomogeneous doping or of a distribution of transition temperatures, which would lead to the strong vortex pinning observed in our magnetization and torque measurements.

Nevertheless, for fields applied parallel to a sheet superconductor Saint-James and De Gennes \cite{degennes} showed that superconductivity can nucleate at the surface and survive the application of fields as large as $H = H_{c3} = 1.695 H_{c2}$. For fields between $H_{c3}$ and $H_{c2}$ there would be a superconducting sheet of thickness $\xi (T)$ while the superconducting order-parameter would be zero in the bulk. Perhaps, such a situation might lead to a flux compression scenario similar to the one invoked for the Wohlleben-effect. However, most of the measurements shown here, as for example, the magnetization data in Figs. 5 and 6 were acquired for fields applied perpendicularly to the c-axis or nearly perpendicular to the superconducting planes. Furthermore, as seen in Fig. 3 and in the phase diagram of Fig. 10, the anomalous hysteresis disappears at $T \sim 11$ K or $T \sim 0.64T_c$. Within the surface superconductivity scenario, flux compression leading to an anomalous magnetic response should be seen all the way up to $T_c$. Therefore, we conclude that a flux compression scenario is not relevant for our experiments.

For granular $d$-wave superconductors one can consider random-junctions \cite{sigrist}, which can also cause paramagnetic Meissner effect. However our sample is not granular, and does not demonstrate signatures for $d$-wave superconductivity at zero or low fields. Therefore, for our observations the  $\pi$-junction scenario is also invalid.

Broken time reversal symmetry within superconductivity was observed in CeCoIn$_5$ at low temperatures and high fields, and was associated with the coexistence of magnetism with a putative FFLO state \cite{kenzelmann}. This would be an intriguing possibility for LiFeAs, which would agree with the claims of Ref. \onlinecite{cho}, although the comparatively much larger area occupied by the anomalous state in the diagram of Fig. 3 (c), points towards an alternative scenario.

Finally, we emphasize that magnetic impurities, dislocations or stoichiometric deficiencies would lead to Curie paramagnetism, which cannot produce any hysteretic response. While a magnetic order-parameter such as ferromagnetism would produce a net hysteretic response having the same sign as the superconducting one. Therefore, the anomalous response seen here \emph{through very distinct experimental techniques} can only be attributed to a novel magnetic response of the superconducting state, produced by the vortices, which carry a net moment along the field (instead of opposing it) and are susceptible to pinning. The net moment along the field can emerge through the induction of chiral pairing components.

Now we argue, how a chiral component may become stabilized inside the mixed phase of a $s_{\pm}$ superconductor. We assume that the low field (below $H_{c1}$) superconducting phase is the fully gapped $s \pm$ state, which is stable against other competing pairing channels. Nevertheless, in the mixed state and close to $H_{c2}$, the Bogoliubov-de Gennes quasiparticles of a generic type II superconductor become gapless at special points in the Brillouin zone associated with the vortex lattice \cite{rasolt}. These gapless quasiparticles can be susceptible to competing pairing channels, which are innocuous for the low field fully gapped state. The effects of competing pairing channels in the mixed phase can also be understood within a phenomenological Landau-Ginzburg theory, see Appendix below. There are gradient couplings among different singlet order-parameters, which are allowed by the symmetry principles. On the other hand, the Zeeman splitting allows for a Lifschitz invariant in the free energy functional \cite{mineev}, which couples triplet \emph{p}-wave order parameters to the gradient of the $s\pm$  order parameter \cite{dutta}. In the presence of such inter-channel couplings, the super-current of the $s \pm$  state can give rise to a singlet $(d_{x^2 - y^2} + id_{xy})$ or a triplet $(p_x + ip_y)$ pairing component in the mixed phase. The induced triplet component is a unitary state with zero spin projection along the magnetic field \cite{dutta}. The chiral components can appear even when the competing channels are repulsive. The competing chiral components carry finite orbital magnetic moment along the external field, which can lead to a paramagnetic response.

In comparison to the 122 Fe-pnictides, in LiFeAs the nesting between electron- and hole-like Fermi surfaces is far more imperfect \cite{borisenko}. This could increase the strength of the $d$-wave fluctuations, when compared to the $s \pm$ pairing channel. ARPES \cite{borisenko} also suggests the presence of shallow hole-pockets and a van Hove singularity at the zone center, which can cause significant ferromagnetic fluctuations, leading to a nonzero coupling constant in the triplet pairing channel. At present it is unclear, which of these competing channels is stabilized. But our results open the possibility of stabilizing a chiral superconducting state in LiFeAs by applying high magnetic fields.


The NHMFL is supported by NSF through NSF-DMR-0084173 and the
State of Florida.  L.~B. is supported by DOE-BES through award DE-SC0002613. \\

\appendix

\section{Ginzburg-Landau formalism}

\subsection{Induction of $d+id$ component}
We first consider the couplings among $A_{1g}$ $s_{\pm}$, $B_{1g}$ $d_{x^2-y^2}$ and $B_{2g}$ $d_{xy}$ pairing channels. The quadratic part of the free energy functional can be written as $f_{quad}=f_{s}+f_{x^2-y^2}+f_{xy}+f_{s,x^2-y^2}+f_{s,xy}+f_{x^2-y^2,xy}$, where
\begin{widetext}
\begin{eqnarray}
f_{s}&=&K_s(D_{j}\psi_{s})^{\ast}D_{j}\psi_s+r_s|\psi_{s}|^2 \\
f_{x^2-y^2}&=&K_{x^2-y^2}(D_{j}\psi_{x^2-y^2})^{\ast}D_{j}\psi_{x^2-y^2}+r_{x^2-y^2}|\psi_{x^2-y^2}|^2 \\
f_{xy}&=&K_{xy}(D_{j}\psi_{xy})^{\ast}D_{j}\psi_{xy}+r_{xy}|\psi_{xy}|^2\\
f_{s,x^2-y^2}&=&K_{s,x^2-y^2}[(D_{x}\psi_{s})^{\ast}D_{x}\psi_{x^2-y^2}-(D_{y}\psi_{s})^{\ast}D_{y}\psi_{x^2-y^2}+c.c.]\\
f_{s,xy}&=&K_{s,xy}[(D_{x}\psi_{s})^{\ast}D_{y}\psi_{xy}+(D_{y}\psi_{s})^{\ast}D_{x}\psi_{xy}+c.c.]\\
f_{x^2-y^2,xy}&=&iK_{x^2-y^2,xy}H[\psi^{\ast}_{x^2-y^2}\psi_{xy}-\psi_{xy}^{\ast}\psi_{x^2-y^2}]
\end{eqnarray}
\end{widetext}
In the above equations $\psi$'s correspond to the order parameters in different channels, $D_{j}=\partial_{j}-i2eA_j$'s are the covariant derivatives and $A_j$'s are the vector potentials in $x$ and $y$ directions. $H$ is the external magnetic field. In addition $K$'s and $r$'s are phenomenological coupling constants. The coupling constants can be chosen in a way to produce $s_{\pm}$ state as the ground state for $H< H_{c1}$. For example, we can assume that only $r_{s}$ changes sign from being positive to negative, as the temperature is lowered below the transition temperature $T_c$, and $r_{x^2-y^2}$, $r_{xy}$ remain positive in the entire temperature range.

The $f_{x^2-y^2, xy}$ term has been considered in the context of high-$T_c$ cuprate superconductors, to obtain a field-induced $id_{xy}$ component in $d_{x^2-y^2}$ wave superconductors \cite{Tesanovic}. This term describes the Zeeman coupling of the external field and the orbital angular momentum of $(d + id)$ superconductor. Such a coupling is always allowed by the symmetry, but the size of this coupling can be very small. For our problem of $s \pm$ superconductor, the primary reason for a field-induced $(d+id)$ component are the combined effects of $s-d$ couplings $K_{s,x^2-y^2}$ , $K_{s,xy}$ and the Landau level structure of the $s \pm$ order-parameter. The presence of $f_{x^2-y^2, xy}$ term only enhances the size of the induced $(d+id)$ component.

Immediately below $H_{c2}$, we can obtain a qualitative understanding of the emergent state by analyzing $f_{quad}$. In the vicinity of $H_{c2}$, the solutions are found from the following Landau-Ginzburg equations
\begin{widetext}
\begin{eqnarray}
[K_s(D_{x}^{2}+D_{y}^{2})+r_s]\psi_s & \approx & 0 \\
K_{s,x^2-y^2}(D_{x}^{2}-D_{y}^{2})\psi_s +r_{x^2-y^2}\psi_{x^2-y^2}+iK_{x^2-y^2,xy}\psi_{xy}& \approx & 0 \\
K_{s,xy}D_{x}D_{y}\psi_s +r_{xy}\psi_{xy}-iK_{x^2-y^2,xy}\psi_{x^2-y^2}& \approx & 0
\end{eqnarray}
\end{widetext}
The $s$-wave order parameter $\psi_s$ is described by the lowest Landau level wave function and in the symmetric gauge $\mathbf{A}=(-Hy/2,Hx/2)$ has the spatial dependence $\psi_s \sim \exp[-(x^2+y^2)/2l^2]$, where $l$ is the magnetic length. The solution for $d$-wave components are described by
\begin{widetext}
\begin{eqnarray}
\psi_{x^2-y^2}&=&\frac{[-r_{xy}K_{s,x^2-y^2}(D_{x^2}-D_{y^2})+iK_{x^2-y^2,xy}HK_{s,xy}D_xD_y]\psi_s}{r_{x^2-y^2}r_{xy}-K_{x^2-y^2,xy}^2h^2}\\
\psi_{xy}&=&\frac{[-r_{x^2-y^2}K_{s,xy}D_xD_y+iK_{x^2-y^2,xy}HK_{s,x^2-y^2}(D_{x}^{2}-D^{2}_{y})]\psi_s}{r_{x^2-y^2}r_{xy}-K_{x^2-y^2,xy}^2H^2}
\end{eqnarray}
\end{widetext}
By using the solution for $\psi_s $, we find $\psi_{x^2-y^2}$ and $\psi_{xy}$ to be respectively real and pure imaginary. Therefore the two $d$-wave components have a relative phase $\pi/2$, and the mixed state acquires a parity and time reversal symmetry breaking $(d+id)$ component, and carries a finite orbital magnetic moment pointing along the external magnetic field. It also becomes clear from the expressions for d-wave components, that the Zeeman coupling of the $(d+id)$ orbital moment plays a secondary role, and causes enhancement of the $d$-wave components. We can also take a Abrikosov vortex lattice solution or a trial disordered vortex lattice solution for $\psi_s$, and still find a corresponding solution for $d+id$ component. According to the above arguments, a $(d+id)$ component can be induced in the mixed state, irrespective of the signs or strengths of $r_{x^2-y^2}$, $r_{xy}$. However, this component will have observable effects, only if $s-d$ mixing terms are sizable, and $r_s$ becomes comparable to $r_{x^2-y^2}$, $r_{xy}$. In LiFeAs there is a lack of nesting among electron and hole pockets, and this may reduce the strength of $s_{\pm}$ coupling, and make the $d$-wave effects relatively stronger.

\subsection{Singlet-triplet mixing and induction of $p+ip$ component}
There can be mixing between singlet and triplet $p$-wave pairing channels due to Zeeman splitting of fermi surfaces, due to a Lifschiz invariant \cite{S2,S3,S4,S5,S6} (we do not consider the mixing due to spin-orbit coupling). We can describe the $\mathbf{d}$ vector of the $p$-wave pairing as $d_{\mu}=d_{\mu j}\hat{x}_{\mu}p_j$, where $\hat{x}$ is a unit vector in the spin space and couples to the Pauli matrices, $p_j$ is the components of relative momentum. We can write the quadratic part of the free energy as $f_{quad}=f_s+f_{p}+f_{s,p}$, where
\begin{widetext}
\begin{eqnarray}
f_{p}&=&K_{p1}(D_j d_{\mu a})^{\ast})(D_j d_{\mu a})+K_{p2}(D_b d_{\mu j})^{\ast})(D_j d_{\mu b})+K_{p3}(D_b d_{\mu b})^{\ast})(D_j d_{\mu j})+r_pd_{\mu j}^{\ast}d_{\mu j} \nonumber \\
& &-ig_1\epsilon_{\alpha \mu \nu} H_{\alpha} d_{\mu j}^{\ast} d_{\nu j}+g_2H_{\mu}H_{\nu}d_{\mu j}^{\ast}d_{\nu j}\\
f_{s,p}&=&-ig_3 H_{\mu} d_{\mu j}^{\ast} D_j\psi_s+c.c
\end{eqnarray}
\end{widetext}
where $K$'s, $r_p$, $g_1$, $g_2$ and $g_3$ are phenomenological coupling constants. The $f_{s,p}$ term is the Zeeman splitting induced Lifshitz coupling among $p$-wave and $s$-wave pairing order parameters, and $g_1$ term describes the coupling between the spin moment of the Cooper pair and the external field. In the context of A-phase of $^3He$ in the external field \cite{S7}, and also in certain ferromagnetic triplet superconductors, $g_1$ term leads to a non-unitary $p$-wave pairing. If we consider $s$-wave to be the dominant pairing in the low field limit, we can solve for the following approximate Landau-Ginzburg equations in the vicinity of $H_{c2}$,
\begin{widetext}
\begin{eqnarray}
[K_s(D_{x}^{2}+D_{y}^{2})+r_s]\psi_s & \approx & 0 \\
-ig_3H_{\mu}D_j\psi_s+r_{p}d_{\mu j}-ig_1\epsilon_{\mu \alpha \beta}H_{\alpha}d_{\beta j}+g_2H_{\mu}H_{\nu}d_{\nu j}&=&0
\end{eqnarray}
\end{widetext}
If we choose the field along $z$ direction, and $\psi \sim \exp[-(x^2+y^2)/2l^2]$, only nonzero p-wave components are $d_{3j}$, and
\begin{equation}
d_{3j}=\frac{i g_3 H D_j \psi_s}{r_p+g_2H^2}
\end{equation}
and leads to a unitary, $S_z=0$, chiral $(p_x+ip_y)$ pairing. In the tilted field, the equations become more cumbersome, but we always find a chiral $S_z=0$ paired state. The effects of such chiral $p$-wave component can be observable only for a sizable $g_3 H$ and comparable $s$-wave and $p$-wave couplings. In general disorder induced broadening effects can diminish the inter-channel couplings, and therefore the emergent chiral state will be fragile against disorder effects.

\end{document}